\documentstyle{mn}

\newif\ifAMStwofonts

\def\Hip{{\it Hipparcos\/}}
\def\Msun{\ifmmode{~M_\odot}\else$M_\odot$~\fi}
\def\Mden{\ifmmode{~M_\odot $pc$^{-3}}\else$M_\odot $pc$^{-3}$\fi}
\def\kms{\ifmmode{$~km\thinspace s$^{-1}}\else km\thinspace s$^{-1}$\fi}

\def\Msun{\ifmmode{~{\rm M}_\odot}\else${\rm M}_\odot$~\fi}
\def\kms{\ifmmode{$~km\thinspace s$^{-1}}\else km\thinspace s$^{-1}$\fi}
\def\etal{\ifmmode{${\it et al.}$}\else {\it et al.}\fi}
\def\Mden{\ifmmode{~{\rm M}_\odot~{\rm pc}^{-3}}
 \else${\rm M}_\odot$~pc$^{-3}$\fi}
\def\la{\mathrel{\mathchoice {\vcenter{\offinterlineskip\halign{\hfil
$\displaystyle##$\hfil\cr<\cr\noalign{\vskip1.5pt}\sim\cr}}}
  {\vcenter{\offinterlineskip\halign{\hfil$\textstyle##$\hfil\cr<\cr
  \noalign{\vskip1.0pt}\sim\cr}}}
  {\vcenter{\offinterlineskip\halign{\hfil$\scriptstyle##$\hfil\cr<\cr
  \noalign{\vskip0.5pt}\sim\cr}}}
  {\vcenter{\offinterlineskip\halign{\hfil$\scriptscriptstyle##$\hfil
  \cr<\cr\noalign{\vskip0.5pt}\sim\cr}}}}}

\def\bea{\begin{array}}
\def\eea{\end{array}}
\def\beq{\begin{equation}}%
\def\eeq{\end{equation}}
\def\ben{\begin{eqnarray}}
\def\een{\end{eqnarray}}

\renewcommand{\b}[1]{\bmath{#1}}

\newcommand{\bv} {\b{v}}
\newcommand{\be} {\b{e}}
\newcommand{\bV} {\b{V}}

\newcommand{\bu} {\b{u}}
\newcommand{\bI} {\b{I}}

\def\spose#1{\hbox to 0pt{#1\hss}}
\def\lta{\mathrel{\spose{\lower 3pt\hbox{$\mathchar"218$}}
     \raise 2.0pt\hbox{$\mathchar"13C$}}}
\def\gta{\mathrel{\spose{\lower 3pt\hbox{$\mathchar"218$}}
     \raise 2.0pt\hbox{$\mathchar"13E$}}}

\def\kms{\mbox{$\,{\rm km}\,{\rm s}^{-1}$}}
\def\kpc{\mbox{$\,{\rm kpc}$}}
\def\parc{\mbox{$\,{\rm pc}$}}
\def\Msun{\mbox{$\,{\rm M}_\odot$}}

\ifoldfss
  \ifCUPmtlplainloaded \else
    \NewTextAlphabet{textbfit} {cmbxti10} {}
    \NewTextAlphabet{textbfss} {cmssbx10} {}
    \NewMathAlphabet{mathbfit} {cmbxti10} {} 
    \NewMathAlphabet{mathbfss} {cmssbx10} {} 
  \fi
  \ifAMStwofonts
    \ifCUPmtlplainloaded \else
      \NewSymbolFont{upmath} {eurm10}
      \NewSymbolFont{AMSa} {msam10}
      \NewMathSymbol{\upi}     {0}{upmath}{19}
      \NewMathSymbol{\umu}     {0}{upmath}{16}
      \NewMathSymbol{\upartial}{0}{upmath}{40}
      \NewMathSymbol{\leqslant}{3}{AMSa}{36}
      \NewMathSymbol{\geqslant}{3}{AMSa}{3E}

      \let\leq=\leqslant 
       
    \fi
  \fi
\fi 

\ifnfssone
  \newmathalphabet{\mathit}
  \addtoversion{normal}{\mathit}{cmr}{m}{it}
  \addtoversion{bold}{\mathit}{cmr}{bx}{it}
  \def\textbfit{\protect\txtbfit}
  
  \long\def\txtbfit#1{{\fontfamily{cmr}\fontseries{bx}\fontshape{it}%
    \selectfont #1}}

  \newmathalphabet{\mathbfit} 
  \addtoversion{normal}{\mathbfit}{cmr}{bx}{it}
  \addtoversion{bold}{\mathbfit}{cmr}{bx}{it}
  \newmathalphabet{\mathbfss} 
  \addtoversion{normal}{\mathbfss}{cmss}{bx}{n}
  \addtoversion{bold}{\mathbfss}{cmss}{bx}{n}
  \ifAMStwofonts
    \ifCUPmtlplainloaded \else
      \UseAMStwoboldmath
      \makeatletter
      \new@mathgroup\upmath@group
      \define@mathgroup\mv@normal\upmath@group{eur}{m}{n}
      \define@mathgroup\mv@bold\upmath@group{eur}{b}{n}
      \edef\UPM{\hexnumber\upmath@group}
      \new@mathgroup\amsa@group
      \define@mathgroup\mv@normal\amsa@group{msa}{m}{n}
      \define@mathgroup\mv@bold\amsa@group{msa}{m}{n}
      \edef\AMSa{\hexnumber\amsa@group}
      \makeatother
      \mathchardef\upi="0\UPM19
      \mathchardef\umu="0\UPM16
      \mathchardef\upartial="0\UPM40
      \mathchardef\leqslant="3\AMSa36
      \mathchardef\geqslant="3\AMSa3E

      \let\leq=\leqslant 

    \fi
  \fi
\fi 

\ifnfsstwo
  \def\textbfit{\protect\txtbfit}
  
  \long\def\txtbfit#1{{\fontfamily{cmr}\fontseries{bx}\fontshape{it}%
    \selectfont #1}}

  \DeclareMathAlphabet{\mathbfit}{OT1}{cmr}{bx}{it}
  \SetMathAlphabet\mathbfit{bold}{OT1}{cmr}{bx}{it}
  \DeclareMathAlphabet{\mathbfss}{OT1}{cmss}{bx}{n}
 \SetMathAlphabet\mathbfss{bold}{OT1}{cmss}{bx}{n}
  \ifAMStwofonts
    \ifCUPmtlplainloaded \else
      \DeclareSymbolFont{UPM}{U}{eur}{m}{n}
      \SetSymbolFont{UPM}{bold}{U}{eur}{b}{n}
      \DeclareSymbolFont{AMSa}{U}{msa}{m}{n}
      \DeclareMathSymbol{\upi}{0}{UPM}{"19}
      \DeclareMathSymbol{\umu}{0}{UPM}{"16}
      \DeclareMathSymbol{\upartial}{0}{UPM}{"40}
      \DeclareMathSymbol{\leqslant}{3}{AMSa}{"36}
      \DeclareMathSymbol{\geqslant}{3}{AMSa}{"3E}

      \let\leq=\leqslant 

    \fi
  \fi
\fi 

\ifCUPmtlplainloaded \else
  \ifAMStwofonts \else 
    \def\upi{\pi}
    \def\umu{\mu}
    \def\upartial{\partial}
  \fi
\fi

\title[The local density of matter mapped by \Hip]
      {The local density of matter mapped by \textbfit{Hipparcos}}

\author[J. Holmberg and C. Flynn]
{Johan Holmberg$^1$ and Chris Flynn$^2$ \\
$^{1}$ Lund Observatory, Box 43, SE-22100 Lund, Sweden\\
$^{2}$ Tuorla Observatory, V\"ais\"al\"antie 20, FI-21500, Piikki\"o, Finland}

\date{Accepted 1999 June 16. Received 1999 May 19; in original form 1998
  December 30}

\begin{document}
\maketitle

\begin{abstract}

  We determine the velocity distribution and space density of a volume complete
sample of A and F stars, using parallaxes and proper motions from the \Hip\
satellite. We use these data to solve for the gravitational potential
vertically in the local Galactic disc, by comparing the \Hip\ measured space
density with predictions from various disc models.  We derive an estimate of
the local dynamical mass density of $0.102 \pm 0.010 \Msun\parc^{-3}$, which
may be compared to an estimate of $0.095 \Msun\parc^{-3}$ in visible disc
matter.  Our estimate is found to be in reasonable agreement with other
estimates by Cr\'ez\'e et al. (1998) and Pham (1997), also based on \Hip\ data.
We conclude that there is no compelling evidence for significant amounts of
dark matter in the disc.
\end{abstract}


\section{Introduction}

  The European Space Agency's \Hip\ satellite has produced very accurate
distances and proper motions for a complete set of nearby stars. These data
permit a reassessment of the amount of matter in the local Galactic disc
because of two significant improvements. Firstly, the kinematics and vertical
density distribution of stars used to trace the disc potential can be
determined with much higher accuracy, permitting a better determination of the
total amount of gravitating matter in the local disc. Secondly, \Hip\ improves
the measurement of the local Luminosity Function, so that the amount of matter
in the disc in visible components can be better estimated.  Both these
improvements lead to a better evaluation of any difference between the amount
of visible disc matter (in gas and dust, stars, stellar remnants and sub-stellar
objects) and the total amount of gravitating matter.

  Any difference between these quantities would imply that there remain mass
components in the disc which have not yet been directly observed, i.e. disc
dark matter.  The first proposal that there might be a lot of such unobserved
matter dates to Oort (1932, 1960).  Oort used the now classical method of
solving the combined Poisson-Boltzmann equations for the kinematical and
density distribution of a population of stars, assumed to be stationary in the
total matter distribution of the disc.  Oort found that approximately one third
of the local disc mass remained unaccounted for.  Modern studies can arguably
be dated from the work of Bahcall (1984a,b,c) who introduced a new method of
describing the visible disc matter as a series of isothermal components, in a
reanalysis and corroboration of Oort's results.

  Studies since 1984 fall into two types, with mildly conflicting results.
Firstly, the measurement of the kinematics and vertical density falloff in the
disc of a stellar tracer (with the important improvement that both are
determined from a single sample) from which the local volume density and/or
local column density of matter in the disc may be derived. Kuijken \& Gilmore
(1989a,b,c), Kuijken (1991) found little evidence for disc dark matter using
faint K dwarfs at the South Galactic Pole (SGP), whereas Bahcall, Flynn \&
Gould (1992), using bright K giants at the SGP, confirmed Bahcall's (1984)
claim of dynamically significant disc dark matter (over 50 per cent by mass)
although with weakened statistical significance. The inclusion of the
kinematics and local density of nearby K giants by Flynn \& Fuchs (1994) to the
sample of Bahcall et al. (1992) reduced the discrepancy between visible and
total matter; they derived best fits with only 20 per cent of the disc mass in
dark form. Secondly, general star counts in combination with mass models of the
Galaxy may be used to infer the amount of disc dark matter. Bienaym\'e, Robin
\& Cr\'ez\'e (1987), in a study of star counts to faint magnitudes, found that
the data could be well fitted by considering the mass distribution of the known
matter only.

  Very recently Cr\'ez\'e et al. (1998) have used the results from the \Hip\
mission to measure the kinematics and vertical falloff of complete samples of
nearby A-F stars. The determination of the local density of gravitating matter
then proceeds via the Poisson-Boltzmann equations for the density of a tracer
component in a self-gravitating disc. Only for A and F type stars does the
density change sufficiently with vertical distance $z$ from the disc mid-plane
that the total local density of matter can be measured.  The very great
improvement comes because the distances to the tracer stars are measured
directly for a complete sample via parallaxes, rather than via photometric
methods as has always been the case before \Hip.\ They find that no dark matter
is required in the local disc, measuring the total amount of gravitating matter
as $\rho_0 = 0.076 \pm 0.015 \Msun\parc^{-3}$ while the amount of visible
matter is estimated to lie in the range $\rho_{\rm vis} = [0.06, 0.10]$ \Mden.

  Another study using \Hip\ data was made by Pham (1997), who used a different
method to study F stars, where he determined scale heights and velocity
dispersions for his chosen sample and then, assuming isothermality, used the $2
\pi G h_{z}^{2} \rho_0 = \sigma_{z}^{2}$ relation. He finds the local mass
density to be $\rho_0 = 0.11 \pm 0.01 \Msun\parc^{-3}$.

  In this paper we have also used A-F stars in the \Hip\ catalogue to measure
the local density of matter $\rho_0$, using an analysis which differs somewhat
from that of Cr\'ez\'e et al. (1998) and Pham (1997). We confirm the Cr\'ez\'e
et al. (1998) and Pham (1997) result of little significant disc dark matter. We
use the method of von Hoerner (1960) as described by Flynn \& Fuchs (1994), in
which the velocity distribution of a tracer population at the mid-plane may be
integrated in a model of the local disc potential to yield its density falloff
in the vertical direction. The predicted falloff may then be compared with the
\Hip\ measured falloff, and the disc model evaluated.  As is usual, our disc
models describe the density stratification of known and putative dark
components in the disc.

 In section 2 we describe the selection of the sample from \Hip.\ In section 3
we detail our models of the disc mass stratification, and in section 4 we fit
the data with and without dark matter in the models, finding it unnecessary to
invoke disc dark matter. In section 5 we summarize and conclude.

\section{The \textbfit{Hipparcos} Catalogue and selection of sample}

  In order to measure the local density of matter, we require the vertical
kinematics (i.e. $w_{0}$ velocity distribution at the Galactic midplane,
$f(|w_{0}|)$) and the vertical density law $\nu(z)$ of a suitable tracer
population. \Hip\ parallaxes may be used to estimate distances of tracer stars
up to 200 pc. Only A and F stars develop a useful change in the density
as a function of height $z$ above the plane within 200 pc, so it is to these
stars we confine our study.

  The stars are drawn from the \Hip\ Survey (hereafter the ``Survey''), a
predefined part of the \Hip\ Catalogue intended to be a complete, magnitude
limited stellar sample. For stars of spectral type earlier than or equal to G5
this magnitude limit is $V \leq 7.9 + 1.1~{\rm sin}~|b|$.

  We select the A-F stars as follows.  We select Survey stars with
$-0.2<B-V<0.6$ and apply an absolute magnitude cut of $0.0 < M_V < 2.5$. These
cuts are illustrated in Fig. 1.  This results in 14342 stars. Proper motions
and parallaxes are available for all these stars.  We further divide this
sample into two at an absolute magnitude $M_V = 1.0$. We refer to these two
samples as A and F stars.  The first sample roughly takes in B5 to A5 stars,
while the second covers A0 to F5 as a result of the rather large dispersion in
the absolute magnitude to spectral type relation shown by \Hip\, (Houk et
al. 1997; Jaschek \& G\'omez 1998).

  The vertical velocity dispersion of stars changes quite quickly as a function
of absolute magnitude along the upper main sequence, as a consequence of the
age-velocity relation (Wielen 1977).  For this reason, dividing the sample into
A and F makes the analysis of the velocity-density information in \Hip\
considerably simpler. Furthermore, since the A star sample is likely to be
considerably younger than the F star sample, there is a possibility that the
stars are so young that they are not yet fully mixed into the Galactic
potential, a key assumption in applying the method. By working with an A and an
F star sample we have some chance of detecting such effects. Ideally, one would
prefer to be using old stars, such as dwarfs below the turnoff or K giants. We
cannot use the dwarfs since they are well below the faintness limit of the
Survey part of \Hip.\ The K giants develop only a small change in density with
vertical distance within 200 pc.

\begin{figure}   
\input epsf
\center
\leavevmode
\epsfxsize=0.9
\columnwidth
\epsfbox{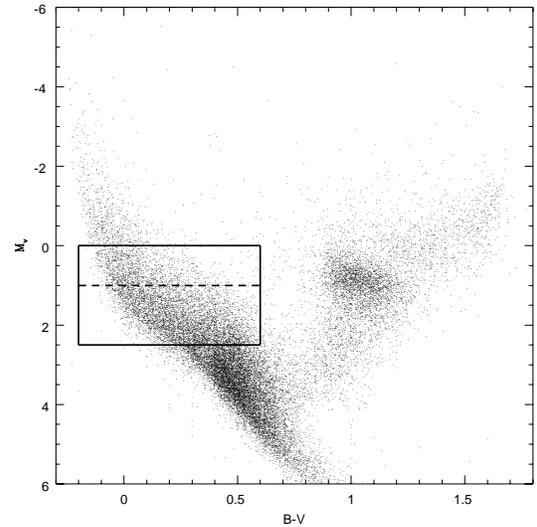}
\caption{Hertzsprung-Russell diagram of \Hip\ survey stars 
with $\sigma_{\pi} / \pi < 15$ per cent. The sample used in this study is 
confined  within the box. The dashed line divide the stars into the
``A''-star (above the line) and ``F''-star sample.}
\end{figure}

  The A star sample contains 4448 stars. We remove stars further than 200 pc,
leaving 2026 stars. The F star sample contains 9894 stars. These stars are
intrinsically fainter than the A stars, and we must apply a 100 pc distance
limit in order to ensure that the sample remains complete. Applying this cut 
leaves 3080 stars.

  For these two samples we use the \Hip\ proper motions and parallaxes to
determine the distribution of vertical velocity and change in density with
height, described in the next two subsections.

\subsection{Stellar density laws}

  The density distribution of the tracer samples as a function of vertical
distance from the galactic plane is determined within a cylinder centered on
the sun. The radius of the cylinder is 100 pc for the F-star sample, and 200 pc
for the A-star sample.  Galactic $(x,y,z)$ coordinates are determined for each
star from the parallax and position on the sky, and the number of stars in 10
pc slices in $z$ is determined. These samples are sufficiently local that the
effects of dust absorption are very minor. We correct for extinction effects
using the extinction model of Hakkila et al. (1997), which is a synthesis of a
number of published studies.

  The number density falloff and mean distance error as a function of vertical
distance from the Sun $|z-z_\odot|$ is shown for the A-stars in the upper
panels of Fig. 2 and for the F-stars in the lower panels. For most of the bins,
the error in the distances to the stars is much less than the bin width,
meaning that corrections to the observed density are small. For the A-stars the
reddening correction has a larger impact than the distance errors, but only
results in a change of the normalization of the counts and does not effect the
slope, which is what we are concerned with here. To illustrate this, the solid
histograms in the left panels show the observed density law in $|z-z_\odot|$,
while the dotted lines show the extinction corrected density law. The
corrections have been calculated using Monte-Carlo methods, in which large
numbers of stars are simulated on the sky using the galactic model of Holmberg,
Flynn \& Lindegren (1997). We simulate the observation of stars by \Hip\ in the
models, including the errors as a function of apparent magnitude and position
on the sky, and this allows us to calculate the small correction between the
observed and true density.

\begin{figure} 
\input epsf
\center
\leavevmode
\epsfxsize=0.9
\columnwidth
\epsfbox{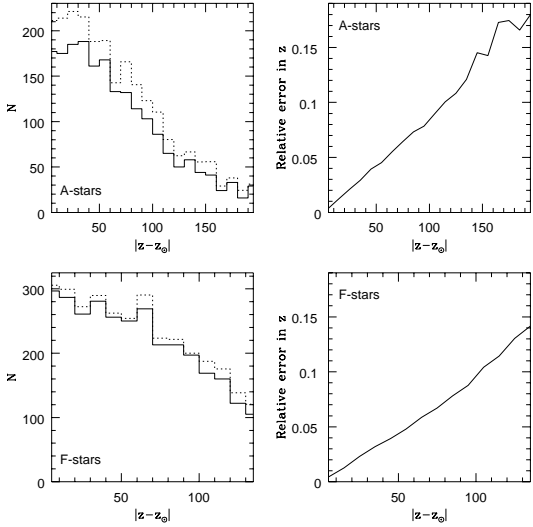}
\caption{Lefthand panels show the number of stars in 10 pc slices in the
samples before (solid) and after (dotted) corrections for distance measurement
errors and reddening for A-stars (upper panel) and F-stars (lower panel).  The
righthand panels show the relative vertical distance error as a function of
distance to the sun, $z-z_\odot$, for the two samples.}
\end{figure}

\subsection{Vertical velocity distribution}  

  \Hip\ gives us for the first time large numbers of accurate parallaxes and
proper motions for a kinematically unbiased sample. Unfortunately, the original
plan to obtain complementary radial velocities measurements was not implemented
although the situation will improve in the near future (Udry et al. 1997).
Full space velocities $(u,v,w)$ are unobtainable for most of the stars, and any
subsample of stars that have radial velocities would be kinematically biased
towards high velocity stars. To get an unbiased estimate of the vertical
velocities $w$ we sacrifice the radial velocity information and work
exclusively with the \Hip\ parallaxes and proper motions.

  We use the standard galactic triad with principal directions
$\hat{\bmath{x}}$ towards the Galactic centre, $\hat{\bmath{y}}$ in the
direction of Galactic rotation, and $\hat{\bmath{z}}$ towards the north
Galactic pole, $\mu_{\ell}$ and $\mu_{b}$ denoting the proper motions in
galactic longitude and latitude in mas, $\pi$ the parallax in mas and
$\kappa$=4.7405 the numerical factor that gives $\bV_{T}$ in km$^{-1}$. The
tangential velocity of a star in the plane of the sky is then defined by

\beq \bV_{T} \equiv{\kappa\over\pi} \left[\bea{r}
          \sin\ell\,\cos b\,\mu_\ell - \cos\ell\,\sin b\,\mu_b \\
          \cos\ell\,\cos b\,\mu_\ell - \sin\ell\,\sin b\,\mu_b \\
                                                 \cos b\,\mu_b
                   \eea \right]
\eeq

The tangential velocity can also be written in terms of the space velocity
$\bv$ (with $\be$ giving the direction to the star) as

\beq
\bV_{T} = \bv - \bv_{R} = \bv - \be v_R = \bv - \be \be^{'}\bv =
(\bI-\bu \bu^{'}) \bv 
\eeq

Combining this we get the equation for the vertical velocity towards the north
Galactic pole

\beq
w = {\kappa \mu_{b} \over \pi \cos b} + u \cos l \tan b + v \sin l \tan b 
\eeq

  The ensemble of tracer stars then gives an estimate of the vertical velocity
distribution function $f(|w_{0}|)$.  Ideally, one would like to determine
$f(|w_{0}|)$ from the proper motions of all the stars in the two samples. We
found it more practical to determine $f(|w_{0}|)$ from stars at low Galactic
latitude, since we use proper motions for the stars and ignore the (incomplete)
radial velocity data for the sample and the velocity distribution function is
dominated by the proper motions rather than the unknown radial velocity at low
Galactic latitude. We chose to use a Galactic latitude cutoff of $|b| < 12^{\rm
o}$. The choice of $12^{\rm o}$ galactic latitude was the best compromise
between obtaining more stars by going to higher latitudes, while keeping $|b|$
low in order to minimize the effect of the unknown radial velocity on
$f(|w_{0}|)$.  

  There are several error sources that affect the velocity distribution
function, $f(|w_{0}|)$. An obvious one is that what we are measuring is not the
distribution at the plane $f(|w_{0}|)$ but the mean of the distribution
$f(|w|)$ within the region delimited by $|b| < 12^{\rm o}$. Since samples taken
further away from the plane are more dominated by kinematically hotter stars,
this effect widens the velocity distribution. Another effect stems from
measurement errors in the parallaxes --- these scatter into the sample more
distant stars. These stars must, in order to meet the apparent magnitude limit
of the survey, be more luminous and hence younger and kinematically colder,
which would narrow the velocity distribution. Finally, we are only measuring
the part of the vertical velocity that comes from the proper motion in
latitude. However, this is the smallest error of the three.  From studies of
the effect on $f(|w_{0}|)$ of varying $|b|$ and the distance limit of the
sample in simulated \Hip\ catalogues we estimate that the combined effect from
these errors are of the order of $0.1\kms$ for the A-sample, and even smaller
for the F-sample, at a cutoff latitude of $|b| < 12^{\rm o}$. For $|b| <
20^{\rm o}$ the correction rises to $0.5\kms$ which would eventually lead to a
20 per cent error in the estimated local density of the disc. For $|b| < 2^{\rm
o}$ the uncertainty in $f(|w_{0}|)$ due to the declining sample size is also
$0.5\kms$. Hence $|b| < 12^{\rm o}$ represents the best compromise between
declining sample size and the corrections described above. The measured
velocity distributions for the two samples are also slightly broadened due to
errors in the parallaxes and proper motions. However, simulations show that
this effect is quite small for our samples. The sample dispersion increases by
a mere $0.04\kms$ for the F-stars, and by $0.16\kms$ for the A-stars due to
these errors.

  The velocity distributions for the two samples, $f(|w_{0}|)$, are shown in
Fig. 3, after a correction for broadening and for the Solar motion relative to
the sample stars of $u=10\kms$,$v=10\kms$ and $w=7\kms$. This Solar motion is
based on an analysis of the the complete survey and is in good
agreement with other determinations based on \Hip\ data (Dehnen \& Binney
1998; Bienaym\'e 1999). There are 723 A-stars and 683 F-stars in these two
distribution functions. The velocity dispersion of the sample corrected for
measurement errors and outliers is 5.7$\pm$0.2 km$^{-1}$ for the A-stars and
8.3$\pm$0.3 km$^{-1}$ for the F-stars.

\begin{figure} 
\input epsf
\center
\leavevmode
\epsfxsize=0.9
\columnwidth
\epsfbox{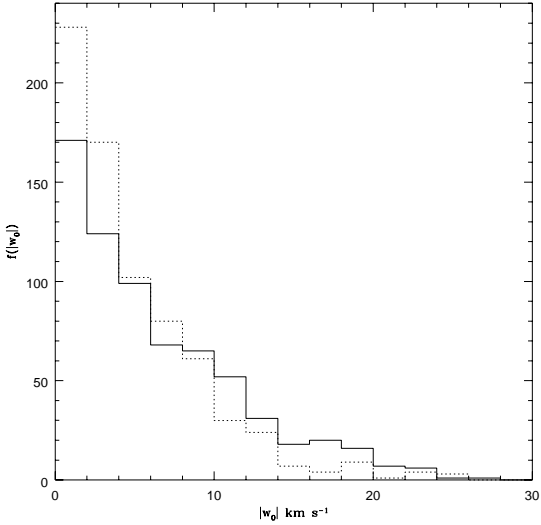}
\caption{Histograms of vertical velocity at the plane, $f|w_0|$, for the A-star
(dotted) and F-star (solid) samples.}
\end{figure}

\begin{figure} 
\input epsf
\center
\leavevmode
\epsfxsize=0.9
\columnwidth
\epsfbox{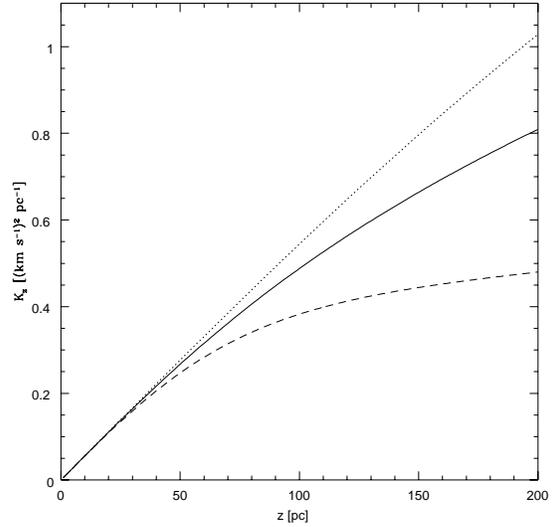}
\caption{Resulting K$_{z}$-force law for three mass models with the same total
midplane density of $\rho_{0} = 0.103 \Msun\parc^{-3}$, of which $0.093
\Msun\parc^{-3}$ is in the disc components and $0.01 \Msun\parc^{-3}$ is in the
dark halo. The solid line shows the reference model from Table 1, the dashed
line model is entirely made up of cold gas ($\sigma_{z} = 4 \kms$) and the
dotted line of old stars ($\sigma_{z} = 20 \kms$).}
\end{figure}

\section{Disc Mass Model}

\begin{table}
\small
\caption{Disc Mass Model}
\begin{center}
\begin{tabular}{rllrrl}
\hline
$i$ &Description&$\rho_i(0)$~~~~~& $\sigma_i$~~~~~  & $\Sigma_i$~~~~~     & Note \\
    &           & \Msun pc$^{-3}$ & km s$^{-1}$ &  \Msun pc$^{-2}$& \\
\hline
1 & H$_2$               & 0.021   ~~~~&   4.0 ~~~~ &   3.0~~~~~~& C   \\
2 & H\thinspace I(1)    & 0.016   ~~~~&   7.0 ~~~~ &   4.0~~~~~~& C   \\
3 & H\thinspace I(2)    & 0.012   ~~~~&   9.0 ~~~~ &   4.0~~~~~~& C   \\
4 & warm gas            & 0.001   ~~~~&  40.0 ~~~~ &   2.0~~~~~~& C   \\
5 & giants              & 0.0006  ~~~~&  17.0 ~~~~ &   0.4~~~~~~& H   \\
6 & $M_V < 2.5$         & 0.0031  ~~~~&   7.5 ~~~~ &   0.9~~~~~~& H   \\
7 & $2.5 < M_V < 3.0$   & 0.0015  ~~~~&  10.5 ~~~~ &   0.6~~~~~~& H   \\
8 & $3.0 < M_V < 4.0$   & 0.0020  ~~~~&  14.0 ~~~~ &   1.1~~~~~~& H   \\
9 & $4.0 < M_V < 5.0$   & 0.0024  ~~~~&  19.5 ~~~~ &   2.0~~~~~~& H   \\
10 & $5.0 < M_V < 8.0$  & 0.0074  ~~~~&  20.0 ~~~~ &   6.5~~~~~~& H   \\
11& $M_V > 8.0$         & 0.014   ~~~~&  20.0 ~~~~ &  12.3~~~~~~& X   \\
12& white dwarfs        & 0.005   ~~~~&  20.0 ~~~~ &   4.4~~~~~~& L   \\
13& brown dwarfs        & 0.008   ~~~~&  20.0 ~~~~ &   6.2~~~~~~& L   \\
14& stellar halo        & 0.0001  ~~~~& 100.0~~~~~ &   0.6~~~~~~& L   \\
\hline 
\end{tabular}
\end{center}
\begin{flushleft}
Notes. C : component constrained by column density \\
H : component constrained by local density using \Hip\ \\
L : component constrained by local density using star counts\\
X : component constrained by column density using HST star counts\\
\end{flushleft}
\end{table}

  We determine the local density of matter by comparing the density law of a
stellar tracer, predicted from it's vertical velocity distribution at the
plane, with the observed density law measured by \Hip. We follow the method of
von Hoerner (1960) in which the distribution of absolute vertical velocity
$f(|w|)$ of the tracer stars at $z=0$, $f(|w_{0}|)$, may be used to construct
the tracer's vertical density profile in the gravitational potential $\Phi$ of
the disc (Flynn \& Fuchs 1994).

  Models for the vertical potential of the Galactic disc are constructed in the
manner of Bahcall (1984a,b,c) in which the local visible components of the disc are
modeled as isothermal components, specified by a local density and a velocity
dispersion. We have made a number of refinements to the mass models of Bahcall
because of observational advances since 1984, in particular for the upper main
sequence, the giant branch, white dwarfs, red dwarfs and brown dwarfs. The
basic model is shown in Table 1, and the resulting K$_{z}$-force law is shown
in Fig. 4 together with two simple one-component models with the same local
density as the reference model.

\subsection{Upper main sequence and giant branch}

For the main sequence above the turnoff and for giants, \Hip\ now provides much
improved measurements of the stellar luminosity function.  Holmberg et
al. (1997) have developed a version of the Bahcall-Soneria Galaxy model with
revised luminosity/colour distributions and scale heights which fits the \Hip\
Survey data.  We have determined the local number density of stars in the
absolute magnitude ranges used by Bahcall, for convenient comparison with his
models. This affects rows 5 to 9 in Table 1. From our model fitting, there are
0.0013 stars pc$^{-3}$ for $M_V < 2.5$, 0.0010 stars pc$^{-3}$ for $2.5 < M_V <
3.0$, 0.0015 stars pc$^{-3}$ for $3.0 < M_V < 4.0$ and 0.0021 stars pc$^{-3}$
for $4.0 < M_V < 5.0$ . For the luminosity range $5.0 < M_V < 8.0$ we take
0.0090 stars pc$^{-3}$ from Jahrei{\ss} \& Wielen (1997). Adopting the
mass-luminosity relation of Henry \& McCarthy (1993), we derive the local mass
densities $\rho_i$ shown in rows 6 to 10 in Table 1.  The local number density
of giants is 0.0005 stars pc$^{-3}$. Adopting a mean mass of 0.9 we derive the
local mass density of giants of 0.0006 \Mden (row 5 of Table 1); Jahrei\ss ~\&
Wielen (1997) derive the same value also from \Hip. While the number densities
of these components of the model can be accurately determined, there is a large
relative error of order 20 per cent in the mass-luminosity conversion in
determining individual $\rho_i$. However, since these stars contribute only
about 10 per cent of the local mass, this error has only a minor impact on the
total mass model. For these stars the velocity dispersion can be estimated
directly using the same technique as for the tracer stars (section 2.2) and are
shown as the $\sigma_i$ in Table 1 in rows 5 to 9. Note that the (\Hip\ based)
observational data essentially constrain the {\em local density} and {\em
velocity dispersion} of these stellar types, and this is emphasised by the H in
column 6 of Table 1.

\subsection{M dwarfs} 

  The column density of disc M dwarfs (Table 1, row 11) can now be measured
directly via star counts using HST (Gould, Flynn \& Bahcall 1998 and references
therein), rather than by extrapolation of the mass function. They measure a
column density of M disc dwarfs of $12.3 \pm 1.8 \Msun\parc^{-2}$.  We
represent the M dwarfs as a single component with a velocity dispersion of $20
\kms$, and we adjust the local density in each model we run to be consistent
with a column density of M dwarfs of $12.3 \Msun\parc^{-2}$.  These stars are
better constrained observationally by column density than by local density, as
indicated by an X in column 6 of Table 1.

\subsection{White dwarfs} 

  Several new estimates of white dwarf (WD) number density have been made since
1984. Oswalt et al. (1996), in an analysis of white dwarfs discovered as proper
motion companions to main sequence dwarfs, after allowing for for the fraction
found in binaries, report the space number density $7.6^{+3.7}_{-0.7} \times
10^{-3}\parc^{-3}$. For an adopted WD mass of 0.6 \Msun\ we have 0.0046
\Mden. Leggett, Ruiz \& Bergeron (1998) find for single WDs in the proper
motion survey of Liebert, Dahn \& Monet (1988) a space density of $3.4 \times
10^{-3}\parc^{-3}$ or 0.002 \Mden. Jahrei{\ss} \& Wielen (1997) report 0.005
\Msun $^{-3}$ for 4 WDs confirmed by \Hip\ to be within 5 pc, very close to the
Wielen (1974) value of 0.007 \Mden, which can be compared with that adopted by
Bahcall (1984), 0.005 \Mden. Knox, Hawkins \& Hambly (1999) finds $4.16 \times
10^{-3}\parc^{-3}$ or 0.0025 \Mden. All these estimates are lower limits, since
they are based on identified white dwarfs. Interestingly Festin (1998) reports
a much higher value of 0.013 \Mden\ for 7 white dwarfs found using the
innovative technique of searching against the opaque screens formed by dark
molecular clouds in Orion and Ophiuchus. Considering the range of values
reported, for the purpose of building mass models we adopt 0.005 \Mden\ but we
will consider model WD densities which cover the range [0.002, 0.013] \Mden.
WDs are primarily constrained by local density measurements as indicated by an
L in column 6 of Table 1.

\subsection{Brown dwarfs} 

  The long-sought freely floating brown dwarfs (BD) of the old disc have now
been found by four groups, namely the Calan-ESO proper motion survey (Ruiz,
Leggett \& Allard 1997), the DENIS mini survey, (Delfosse et al. 1997), a BRI
survey of Irwin, McMahon \& Hazard (1991) and UK Schmidt survey of ESO/SERC
field 287 (Hawkins et al. 1998). Fuchs, Jahrei{\ss} \& Flynn (1998) summarize
the findings of these groups and the implied minimum densities of BDs in the
local disc.  The four surveys lead to density estimates of BDs of 0.46
pc$^{-3}$, $0.076\parc^{-3}$, 0.069 pc$^{-3}$, and 0.15 pc$^{-3}$
respectively. Adopting a BD mass of 0.065 $M_\odot$ this gives mass densities
of 0.03 $M_\odot$ pc$^{-3}$, 0.0049 $M_\odot$ pc$^{-3}$, 0.0045 $M_\odot$
pc$^{-3}$ and 0.01 $M_\odot$ pc$^{-3}$, respectively. These surveys are all in
their early stages and promise an accurate accounting of BDs within a few
years. A fifth survey, 2MASS, has also reported at least 5 BDs in a 420 $\rm
deg^{2}$ survey (Reid et al. 1998) although it is not possible yet to determine
space densities for these BDs. In the model we adopt a BD mass density of 0.008
\Mden, and consider models over the range of [0.004, 0.03] \Mden.  BDs are
primarily constrained by local density measurements as indicated by an L in
column 6 of Table 1.

\subsection{Interstellar matter}

  The remaining component in the model is by far the least well
understood. Many studies and compilations exist in the literature on the
structure and composition on the multi phase ISM (Hollenbach \& Thronson 1987;
Combes 1991) which describe a complex mixture, ranging from cold (10 K) and
dense (50 $M_\odot$ pc$^{-3}$) dark molecular clouds, to the very hot (10$^{6}$
K) and dilute (0.0001 $M_\odot$ pc$^{-3}$) X-ray emitting coronal gas. Here we
adopt the multi-phase model of Bahcall et al. (1992) recognizing that the best
determined parameters are the column density and velocity dispersion, which are
observable quantities, whereas the deduced local density suffers considerably
uncertainties (indicated by a C in column 6 of Table 1).  This four-component
model consists of 3 $ \Msun\parc^{-2}$ at 4 km s$^{-1}$ of molecular gas
(Scoville \& Sanders 1987), two neutral atomic components each of 4 $
\Msun\parc^{-2}$, one cold at 7 km s$^{-1}$ and one warm at 9 km s$^{-1}$ and
finally one hot ionized component with 2 $ \Msun\parc^{-2}$ at 40 km s$^{-1}$
(Kulkarni \& Heiles 1987). The total surface density of these components is
about 13 \Mden, with an uncertainty of about 50 per cent.

\subsection{Adopted velocity dispersions}

  For the early stellar types and giants in the model (rows 5 to 9) the local
velocity dispersion $\sigma_i$ is rather accurate because they are derived from
\Hip\ data. For all other non-gaseous components the velocity dispersion is not
so well determined, and we simply adopt a uniform value of 20 $\kms$ for what
amounts to K to M dwarfs, white dwarfs and brown dwarfs.  We are concerned in
this paper with using the \Hip\ data to measure the local density of matter,
whereas the exact choice of $\sigma_i$ for these components only affects the
column density of the models. To illustrate that the adopted value of 20 \kms
is consistent with known constraints, we plot in Fig. 5 the velocity
distribution of the combined stellar components in the model (rows 5 to 12 in
Table 1) and an observational determination of $f(|w_{0}|)$ by Fuchs \&
Jahrei{\ss} (private communication) based on \Hip\ data for nearby stars.
For purposes of determining the local density the match between model and data
is suitably accurate.

\begin{figure} 
\input epsf
\center
\leavevmode
\epsfxsize=0.9
\columnwidth
\epsfbox{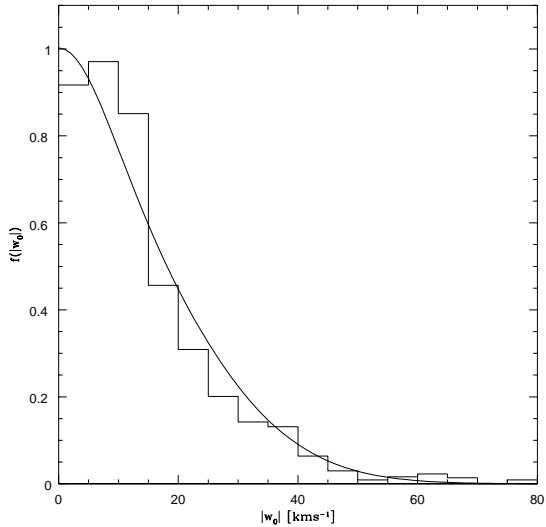}
\caption{The velocity distribution function for the stellar components of our
model (solid line) compared to that observed in the solar neighbourhood by
Fuchs \& Jahrei{\ss} (private communication, histogram).}
\end{figure}

\section{Model fitting and results}

  We now use the velocity distribution functions, $f(|w_{0}|)$, as determined
in section 2.2 for the A- and F-star samples, to predict the density falloff in
$z$ of such stars in our disc mass model.  The density falloff $\nu(z)$ is
determined via the relation (Fuchs \& Wielen 1993; Flynn \& Fuchs 1994):

\begin{equation}
                \nu(z) = 2 \int_{\sqrt{2\Phi}}^\infty 
                { f(|w_0|)~ w_0~{\rm d}w_0 \over \sqrt{w_0^2-2\Phi} }
\end{equation}

where $w_0$ is the vertical velocity at the mid-plane, $z=0$ and $\Phi(z)$ is
the total gravitational potential generated by the mass model.
Equation 4 ignores the radial term in Poisson's equation. If the term
is included, Poisson's equation takes the form (Binney \& Merrifield 1998):

\begin{equation}
  4\pi G \rho = -{\partial K_{z} \over \partial z}
   + 2(B^2-A^2) 
\end{equation}

  where $A$ and $B$ are the Oort constants. Using the values derived by Feast
\& Whitelock (1997), $A=14.82\kms\kpc^{-1}$ and $B=-12.37\kms\kpc^{-1}$ the
correction term amounts to $-0.0025 \Msun\parc^{-3}$. However, in a new
determination of Mignard (1998) quoted in Kovalevsky (1998), the values are $A
= 11.7$\kms\kpc$^{-1}$ and $B = -10.5$\kms\kpc$^{-1}$ giving a smaller
correction of $-0.001 \Msun\parc^{-3}$. These corrections are much smaller than
other sources of error, and can be safely neglected.

  We integrate Eqn. 4 numerically using the model in Table 1 and the two
velocity distribution functions $f(|w_{0}|)$ shown in Fig. 3 for the A- and
F-stars. The integration is performed by using a velocity interval of $\Delta
w_0 = 0.01 \kms$ in order to produce a fairly smooth number density falloff ---
a too large velocity interval (e.g. 0.1 $\kms$) causes a visible and
non-physical sawtooth effect in the the resulting number density law $\nu$.

\subsection{A-star sample}

  The number density falloff for the A-stars resulting from the mass model in
Table 1 is shown by the dotted line in Fig. 6, versus the \Hip\ data
(histogram) and is already seen to be a good fit. We attempted to improve the
fit by adding or removing mass from the model, and minimising the difference
between the predicted and observed number density as a function of $z$ using a
standard $\chi^{2}$ statistic.  It made little difference to the fitting
whether this mass was removed proportionally from all rows in the model or
(more reasonably) from the most uncertain rows in the model, as one expects
since the measurement is so local.  For the A-stars the resulting best fitting
local density was $\rho_{0}=0.103 \pm 0.006 \Msun\parc^{-3}$. i.e. just the
same as the reference model of Table 1.

\begin{figure} 
\input epsf
\center
\leavevmode
\epsfxsize=0.9
\columnwidth
\epsfbox{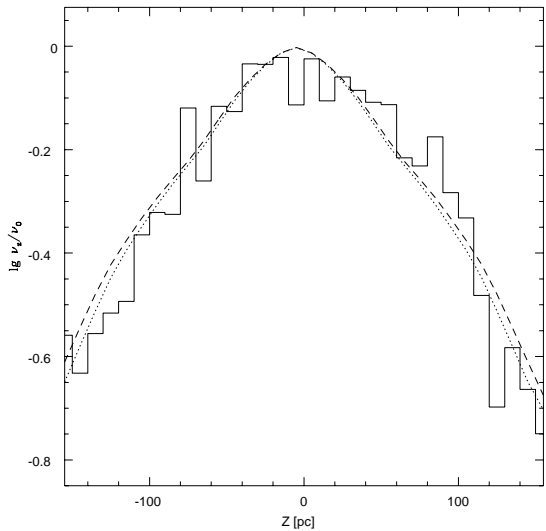}
\caption{The vertical distribution of A-stars (histogram) versus the prediction
of the best fitting mass-model with $\rho_{0} = 0.103 \Msun\parc^{-3}$ (dotted
line). Also shown is the F-star model with $\rho_{0} = 0.094 \Msun\parc^{-3}$
(dashed line).}
\end{figure}

\subsection{F-star sample}

  We performed a similiar fitting for the F-stars as for the A-stars.  For the
F stars the best fitting vertical density distribution is shown by the dashed
line in Fig. 7, and corresponds to a local density of $\rho_{0}=0.094 \pm 0.017
\Msun\parc^{-3}$. This best fitting model has slightly less local matter 
(by approximately 9 per cent) than the basic model in Table 1. As for the
A-stars, it made little difference whether the mass was removed in proportion
from the whole model or from specific rows. To make this best fitting model
mass was removed proportionally from the gas and white/brown dwarf components.

\begin{figure}  
\input epsf
\center
\leavevmode
\epsfxsize=0.9
\columnwidth
\epsfbox{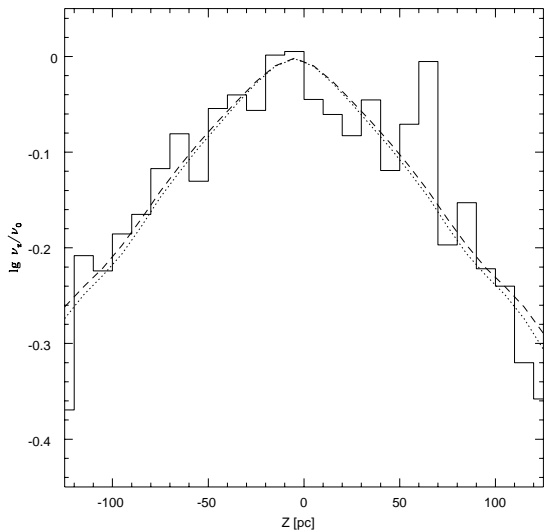}
\caption{The vertical distribution of F-stars (histogram) versus the prediction
of the best fitting mass-model with $\rho_{0} = 0.094 \Msun\parc^{-3}$ (dashed
line). Also shown is the A-star model with $\rho_{0} = 0.103 \Msun\parc^{-3}$
(dotted line)}
\end{figure}

\subsection{Error estimates}

  We have estimated our confidence limits via a series of Monte Carlo
simulations of observations drawn from synthetic \Hip\ survey catalogues. The
catalogues were created from a kinematical Galactic model constructed to fit
the \Hip\ and Tycho catalogues (Holmberg 1999). In each simulation the number
density and velocity distribution of the tracer stars was exactly known. A
sample was drawn from the simulation taking into account the \Hip\ magnitude
limits and magnitude dependent parallax and proper motion errors. The local
density of the disc was then determined from the artificial sample in the same
manner as for the actual data, by measuring $\nu(z)$ and $f(|w_0|)$ and fitting
mass models. Several thousand Monte-Carlo simulations of this type allowed us
to estimate our confidence limits for a given disc density.

  In the modeling of the A-star sample the velocity dispersion was $5.72 \pm
0.18\kms$ and the best fit density is $\rho_{0}$=0.101 $\pm$ 0.006
\Msun\parc$^{-3}$, with a 95 per cent confidence limit of $\pm$ 0.011 \Msun
\parc$^{-3}$, showing that the uncertainty in the density is totally dominated
by the quadratic dependence of the velocity dispersion.  The F-star sample had
a velocity dispersion of $8.14 \pm 0.24\kms$ with a best fit density of
$\rho_{0}$ = 0.103 $\pm$ 0.017 \Msun\parc$^{-3}$, with a 95 per cent confidence
limit of $\pm$ 0.023 \Msun \parc$^{-3}$. This shows a differing effect in which
the variance of the resulting density is dominated by the uncertainty in the
determination of the fall-off of the tracer density.

 In the $\chi^{2}$ fitting process the density distribution resulting from the
mass model is considered to be a distribution function without errors. This has
the effect that the resulting value of $\chi^{2}$ is increased by the errors
from the velocity distribution function of the tracer stars. In the simulation
of the complete sample the mean value of $\chi^{2}$ was 59.7 instead of the
expected 56.3 for a model without errors. For the actual \Hip\ samples the
resulting values of $\chi^{2}$ was 77.76 for 57 DF in the total sample, 44.59
for 31 DF in the A-star sample and 33.11 for 25 DF in the F-star sample.

\subsection{Disc column density}

  The total column of gravitating matter in the mass models that best fits all
our samples is quite close to the $48 \Msun \parc^{-2}$ of visible matter in
our basic disc mass model of Table 1. We strongly emphasize here that our
samples are so local that we cannot put any constraints on the column, since
most of it is above our tracers. An illustration of this is that if the
velocity dispersion of the hot disc stars in the mass model is increased from
$20\kms$ to $25\kms$ the column inflates by $7 \Msun \parc^{-2}$ but the local
density estimate changes with only $0.002 \Msun\parc^{-3}$ (see also section
3.6 above).

\subsection{Dependence on disc mass model}

  How critical is the formulation of the mass model to the result? We tested
this by using the quite different model from Flynn \& Fuchs (1994), slightly
refined by the inclusion of \Hip\ observations (Fuchs 1998, private
communication). This model is quite different in that rather than modeling
individual mass components by isothermals, as we do here, Flynn \& Fuchs
reconstructed the disc potential $\Phi$ using the velocity distribution
function of K and M dwarfs, to which a small component of young stars and gas
had been added.  Adopting this model for $\Phi$ resulted in a change of only
$0.003 \Msun\parc^{-3}$ to our best-fitting local mass density determinations.

\subsection{Combined A- and F-star sample}

  Our conclusion in this section is that the dynamically estimated mass in the
solar neighbourhood is completely accounted for by the identified material in
gas, stars and stellar remnants without any need for dark matter in the disc.
Since both the A- and F-star samples agree on the local density within their
errors, we combined the sample and derived a best fitting local density of
$\rho_{0}$=0.102 $\pm$ 0.006 \Msun\parc$^{-3}$, with 95 per cent confidence
limits of $\pm$ 0.010 \Msun \parc$^{-3}$.

\subsection{Comparison to other work}

  Our estimate of the local mass density is $\rho_0 = 0.102 \pm 0.006
\Msun\parc^{-3}$. From very similar \Hip\ data Pham (1997) estimates $\rho_0 =
0.11 \pm 0.01 \Msun\parc^{-3}$ and Cr\'ez\'e et al. (1998) estimate $\rho_0 =
0.076 \pm 0.015 \Msun\parc^{-3}$. These estimates differ at less than the
2$-\sigma$ level, but we have investigated possible causes for the discrepancy
between them.  The major difference between the determinations is the method
applied to the data.  Pham (1997) used the $2 \pi G h_{z}^{2} \rho_0 =
\sigma_{z}^{2}$ relation, which implicitly assumes that the tracer stars
(i.e. the A- and F-stars) have a Gaussian velocity distribution. Fig. 3 clearly
illustrates that this is in fact only an approximation. Had we carried out our
analysis using a Gaussian distribution function rather than the measured
distribution function $f(|w_{0}|)$ we would have recovered a local density of
$\rho_0 = 0.110\Msun\parc^{-3}$.  Cr\'ez\'e et al. (1998) estimate the local
density by representing the vertical acceleration in the disc $K_z$ by a linear
function, $K_z = az$, where $a$ is a constant. $K_z$ may not actually be linear
in the region of interest ($z \la 100$ pc).  As can be seen in Fig. 4., $K_z$
would not be well represented by a linear function in the model adopted here
for all known mass components of the disc (solid line).  If we had assumed a
strictly linear vertical force law, then tests showed that our best fitting
local density would have been reduced to $0.089\Msun\parc^{-3}$. This effect
could explain part of the difference between our result and Cr\'ez\'e et
al. (1998).  In summary, the small differences between the three local density
estimates using \Hip\ data are probably a result of the three different methods
used.

\begin{table}
\small
\caption{Local density estimates}
\begin{center}
\begin{tabular}{lll}
\hline
Density $\rho_0$& Error  & Reference \\
(\Msun pc$^{-3}$) &(\Msun pc$^{-3}$)  &\\
\hline
0.185 & 0.020 & Bahcall (1984b) \\
0.210 & 0.090 & Bahcall (1984c) \\
0.105 & 0.015 & Bienaym\'e, Robin \& Cr\'ez\'e (1987) \\
0.260 & 0.150 & Bahcall et al. (1992) \\
0.110 & 0.010 & Pham (1997) \\
0.076 & 0.015 & Cr\'ez\'e et al. (1998) \\
0.102 & 0.010 & This paper \\
\\ 
0.150 & 0.026 & Straight average \\
0.108 & 0.011 & Variance weighted average \\
\hline
\end{tabular}
\end{center}
\end{table}

  Stothers (1998) gives a detailed summary of measurements of the local disc
density over more than 60 years since Oort (1932). Stothers draws attention to
the fact that there are sufficiently many determinations of $\rho_0$ that one
might apply the central limit theorem and derive the best density from the
straight average, which is $0.141\pm.0.010\Msun\parc^{-3}$.  However, in
contrast to most older studies, almost all determinations from the last 20
years have clearly stated uncertainty estimates attached to the preferred
density. The data can be found in Table 2 together with our own estimate in
this paper for $\rho_0$. The table shows that high estimates of the density
generally have large uncertainties. The sample mean weighted by the inverse
variance is $0.108\pm.0.011\Msun\parc^{-3}$ which is quite consistent with the
\Hip\ results. One should note that the last three determinations are based on
almost the same data from \Hip. 

\section{Conclusions}

  From a sample of A and F stars taken from the \Hip\ catalogue, we determined
vertical space velocities and density distributions.  Using a method of von
Hoerner (1960) we solved for the vertical potential and hence local density
that would give the observed velocity and density distributions. We find that
local dynamical mass density of the solar neighbourhood is $0.102 \pm 0.010
\Msun \parc^{-3}$ well compatible with the identified material in ordinary
matter leaving no need for dark matter in the disc.  Our estimate is found to
be in reasonable agreement with other estimates, taking errors into account. We
conclude that there is no compelling evidence for significant amounts of dark
matter in the disc.

\section*{Acknowledgments}

  We thank Burkhard Fuchs, Andy Gould and Lennart Lindegren for many
suggestions and comments. We also thank Burkhard Fuchs and Hartmut Jahrei{\ss}
for allowing us to use their data in advance of publication. JH is supported by
the Swedish Space Board. JH also thanks the Finnish Academy for funding several
visits to Tuorla Observatory, where part of this work was carried out. This
paper is based on data from the ESA \Hip\ astrometry satellite.
 
{} 

\end{document}